\documentclass{aastex}
\usepackage{spr-astr-addons}
\usepackage{url}

\RequirePackage{color}

\newcommand{\emaila}{innocent.eya@unn.edu.ng}

\begin{document}

\title{Microglitches in radio pulsars: the role of strange nuggets}
\slugcomment{Not to appear in Nonlearned J., 45.}
\shorttitle{Microglitches in radio pulsars: the role of strange nuggets}
\shortauthors{Eya et al.}

\author{I. O. Eya\altaffilmark{1}} 
\affil{Department of Science Laboratory Technology, University of Nigeria, Nsukka, Nigeria. \\ \emaila{}}

\and \author{E. U. Iyida\altaffilmark{1}}
\and
\author{J. O. Urama\altaffilmark{1}}
\and
\author{A. E. Chukwude\altaffilmark{1}}
\affil{Department of Physics \& Astronomy, University of Nigeria, Nsukka, Nigeria.}
\email{}

\altaffiltext{1}{Astronomy and Astrophysics Research lab, University of Nigeria, Nsukka, Nigeria.}

\begin{abstract}
Strange Nuggets are believed to be among the relics of the early universe. 
They appear as dark matter due to their low charge-to-mass ratio.
Their distribution is believed to be the same as that of dark matter. 
As such, they could be accreted by high magnetic field objects and their collisions with pulsars are inevitable. 
Pulsar glitches are commonly seen as sudden spin-ups in pulsar frequency.
It is still an open debate with regard to mechanisms giving rise to such a phenomenon.
However, there is a class of sudden changes in pulsar spin frequency known as microglitches. 
These event are characterized by sudden small change in pulsar spin frequency ($ \delta \nu/\nu \approx \pm 10^{-9}$).
Clearly, the negative signature seen in some of the events is inconsistent with the known glitch mechanisms.
In this analysis, we suggest that accretion of strange nuggets with pulsars could readily give rise to microglitch events.
The signature of the events depends on the energy of the strange nuggets and line of interaction.
\end{abstract}



\section{Introduction}
\label{sec:intro}
Spinning magnetized neutron stars known as pulsars are one of the most stable natural rotators in the universe.
In some cases their stability rivals that of atomic clock. 
Nonetheless, long term observations have shown a form of abrupt discontinuity in the rotation profile of some pulsars.
Some of these abrupt discontinuity, which manifest as a sudden change in pulsar frequency ($\delta\nu $) are known as glitches  \citep{b12,b6cu,b7,b33}.
Conventionally, pulsar glitches are seen as sudden spin-up (positive change) in pulsar rotation frequency ($\delta\nu$), with fractional size in the range of $ 10^{-11} $ to $ 10^{-5} $.
This spin-up phase, which lasts for about 30 sec \citep{bdod} is usually, (but not always) accompanied by a change in spin-down rate ($\delta \dot{\nu}$).
After the spin-up phase, a relaxation phase normally follows at which the pulsar returns to a pre-glitch frequency or maintains a new steady frequency.

Meanwhile, \cite{b6cu} identified 299 microglitches in 26 pulsars observed at Hartebeesthoek Radio Astronomy Observatory (HartRAO) after 16 years of pulsar timing for each of the pulsars.
These events are characterised by small magnitude change $|\delta\nu/\nu| < 10^{-9} $, which is positive ($\delta\nu/\nu$) in some events and negative ($-\delta\nu/\nu$) in others. 
The negative events could be regarded as anti-glitches.
Earlier than \cite{b6cu}, few microglitches have be reported in rotation-powered pulsars \citep{b6cd,b6cdk}. 
The magnitude of the event though have wide range concentrated within $ \delta\nu/\nu < 10^{-10}$ and $\delta\dot{\nu}/\dot{\nu} < 10^{-3}$.

Pulsar glitch is one of the most promising ways of probing the neutron interior.
After about five decades of pulsar glitch studies, a comprehensive understanding of the nature, origin, signature and recovery of the microglitches and conventional large (micro-) glitches remain not well understood.
In the studies, more attention has been given to conventional glitches than the microglitches.
This one-sided attention may be narrowing the comprehensive understanding of the dynamics of pulsars.
Nonetheless, many models have been proposed toward the understanding of the origins of pulsar glitches (see \cite{b7h}  for a recent review).
Based on the fundamental principles involved, those models could be grouped into two, namely, the starquake models and the angular momentum transfer models.
The starquake models rely on sudden reduction in stellar moment of inertia \citep [e.g.][]{bru,b5a,b43a,b31}.
They could readily account for small size glitches ($\delta\nu/\nu \leq  10^{-7} $)  but they are not convenient for large ones due to the energy involved. 
The angular momentum transfer models rely on sudden transfer of angular momentum from a partially decoupled interior component, which is at a higher velocity than the rest of the pulsar \citep [e.g.][]{b5,b3,b1}.
The angular momentum transfer models for decades now is seen as standards for analysing pulsar glitch events especially the large ones ($ \delta\nu/\nu \geq 10^{-7} $).
Nonetheless, due to recent calculations of the magnitude of neutron star components participating in glitches \citep [e.g.][]{b4a,b6,b7e}, it is now debatable whether the reservoir of the angular momentum involved in glitches is large enough to account for the observed glitch sizes \citep{bb4,b7e1}.
Likewise, the radiative changes accompanying glitches in magnetars and some other high magnetic field pulsars \citep{bwje,bdk,bka,bkou} could not be well understood in the realm of standard glitch models.
In addition to these challenges, the possible free precession in PSRs J1645-0317 and J1830-1059 \citep{b21,b20} is a serious challenge to the idea of vortex pinning in neutron star interior.
If the pinning is strong enough to partially decouple the crustal fluid from the rest of the neutron star, it could have damped the precession in a short time interval than the observed. 
Equally, applying the magnetohydrodynamic coupling of the crust to the core, indicates that the free precession would have decayed in a few hundred of years against the present age of pulsar \citep{b9} . 
Furthermore, the idea that the superfluid neutron vortices in neutron star core coexist with proton flux tube can not give rise to precession \citep{b10,b31}.

Nonetheless in order to reconcile the observed free precession and glitches, \cite{b30} suggested that the way forward is to treat radio pulsars as solid quark stars. 
A solid quark star is a condensed object of quark clusters --- just a rigid body \citep{b31}.
It is different from conventional quarks stars in the sense that the quarks in it could be regarded as fermion gas. 
They are not confined in hadrons of neutrons stars like that of the conventional quark stars. 
This idea of neutron star being a solid quark star is favoured by the observation of massive neutron stars of which the matter in it is best described with very stiff equation-of-state \citep{b8,b8a,b4}.
In this frame, conventional glitches including the large ones being attributed to angular momentum transfer models could, to some extent, be well understood \citep{b31} but not the anti-glitches.

However, there are some physical phenomena that could mimic anti-glitches in timing data. 
Such phenomena include transfer of angular momentum from slower rotating crust to the faster crustal fluid \citep{b14}, curling of magnetic lines \citep{b9l}, accretion of ambient matter \citep{b8k,b18}, strong outburst \citep{b22} and collision of small solid body with neutron star \citep{bhg}. 
Of all these phenomena, none except the collision of small solid body with neutron star could produce a picture of anti-glitch with features of conventional glitch. 
Others appear as a slow deceleration of the pulsar, which could not be regarded as glitch.

On the other hand, \cite{b8b} proposed a glitch mechanism, which is devoid of neutron star vortices.
They treated neutron star as a solid quark star and opined that the accretion of strange nugget (SN) by pulsar could lead to sudden spin-ups known as glitches.
In the mechanism, it is assumed that the distribution of SN in the universe is comparable to that of dark matter. 
With that, they showed that the occurrence rate of small size glitches is in line with the accretion rate of dark matter by stellar objects. 
Strange nuggets are one of the relics of the early universe (for a more detailed review on its formation and nature, see \cite{b9m}). 
If the phase transition that took place at the earliest stage of the universe when the temperature is about 200 MeV is of the first order \citep{b9m,b14}, most of the de-confined quark would be concentrated in the dense invisible SN \citep{b14a,b8b}.
They are baryon in nature but due to their low charge-to-mass ratio, they appear as dark matter \citep{b9m,b8b}.
Their mass ratio is believed to be comparable to that of dark matter \citep{b14a}. 
They could be circulating the galaxy in a similar way as dark matter having higher concentration near the galactic center. 
As such, they could be continuously accreted by high magnetic field objects like pulsars. One physical consequences of such accretion is temporal alteration in dynamics of the pulsar involved --- possible microglitch. 
The incident rate will technically depend on the position of the pulsar in the galaxy.

As at the time of this analysis, over 500 conventional glitches  have been reported in 190 pulsars\footnote{Both in magnetars, rotation-powered pulsars, and accretion powered pulsars} (see JBO and ATNF pulsar glitch catalogue for update).
Most of the events in the catalogues have been extensively studied that the jump in spin parameters are presumed to be associated with a definite signature, $(\delta\nu, \delta\dot{\nu}) = (+,- )$.
However, a positive change ($\delta \dot{\nu} $) have been observed in a few events as seen in conventional glitch database (ATNF and JBO pulsar glitch catalogue).
Meanwhile, recently, anti-glitches of magnitudes, $ \delta\nu \gtrsim 10^{-8} $, was reported in a magnetar \citep{b4b}.
The events were accompanied by hard X-ray burst and decaying softer X-ray emission of which no such event has been reported in rotation-powered pulsar.
The microglitches reported in rotation-powered pulsars,
unlike conventional glitches have combination of signature, ($ \delta\nu,\delta\dot{\nu} $) = ($+,-)$, ($-,-$), ($ +,+ $), ($ -,+ $).
Due to this irregular signature, it becomes difficult to understand microglitch events in the standard glitch theories.

In this analysis, we try to explain both signatures of microglitch events based on a single natural phenomenon by revisiting the accretion of strange nuggets by pulsars.
The results obtained show that interaction of SN with pulsars could readily account for both signature of microglitch events. 
The signature of the glitch depends on the energy of the SN involved.
\section{Sources of data}
The microglitch data is from \cite{b6cu}.
These glitch events are mostly from relatively long period ($ P \geq 0.13 $) pulsars of characteristic age  $ \gtrsim 10^{5} $ yr and magnetic field in the rang $10^{10} < $ B $ \lesssim 10^{12} $.
In the data, 20 of the pulsars exhibit significant fluctuation simultaneously in both spin frequency ($ \nu $) and spin-down rate ($\dot{\nu}$), giving a total of 266 glitches.
With respect to the signature of the glitches, 137 are positive ($ \delta\nu/\nu $),  while 129 are negative ($ -\delta\nu/\nu $).
The contribution of individual pulsars to this number together with other pulsar parameters are summarised in Table 1.
The pulsar  parameters are from the Austrian Telescope National Facility\footnote{http://www.atnf.csiro.au/research/pulsar/psrcat} (ATNF).
To ascertain how the glitches of both sign are distributed, we made the histograms of the distribution of the glitch sizes. 
The outcome is shown in Fig.1.
In all, the range of the microglitch sizes are $ 10^{-11} <|\delta \nu/\nu| \leq 10^{-8} $ with peaks at $ \sim 10^{-9} $.

\begin{table*} 
\centering
\caption{
Parameters of the HartRAO pulsars}
   \begin{center} 
  \begin{tabular}{@{}|lcrrccccc|@{}}
  \hline
Pulsar & Ng  & Ng  &P  &	$ \dot{\nu}\times10^{-15} $	& $ \tau_{c} $ &$\dot{R}_{g}^{+ve} $ &$\dot{R}_{g}^{-ve} $ & r \\
J Name & +ve & -ve &(s)&($  Hzs^{-1}$)	 & (10$ ^{5} $yr) & ($ yr^{-1} $)& ($ yr^{-1} $) & (kpc)\\
       
\hline 
0738-4042	& 9 & 2  &	0.37&   -11.50 	 &	36.8 &0.21&   0.16	&9.1\\
0742-2822	& 24 & 22&  0.17&   -604.87 &	1.6 &	1.50&  0.55 &9.5\\
0837-4135	& 2 & 2 &	0.75&	-6.27	&	33.6 &	0.21&	0.21&8.9\\
1001-5507	& 5  & 3 &  1.44&	-24.90    &  4.4 &	0.32&	0.19&8.5\\
1056-6258	& 10 & 4 &	0.42&	-20.04		&18.7&	0.72&	0.3&8.0\\
1224-6407	& 5 & 2 &	0.22&	-105.70		&6.9 &	0.49&	0.19&7.4\\
1243-6423	& 3 & 6 &	0.39&	-29.82		 &13.7&	0.25&	0.50&7.6\\
1326-5859 	& 7 & 10 &	0.48&	-14.17		 &23.4&	0.47&	0.68&7.1 \\
1327-6222	& 13 & 18 & 0.53&	-67.23   &	4.5 &	0.56&	0.50&6.9\\
1359-6038	& 7 & 4 &	0.13&	-389.91		 &3.2&	0.62&	0.35&6.4\\
1401-6357	& 9 & 4 &	0.84&	-23.55 &	8.0 &	0.65&	0.15&7.5\\
1453-6413	& 3 & 1 &	0.18&	-85.24	 &	10.4 &	0.24&	0.26&6.8\\
1559-4438	& 2 & 1 &	0.26&	-15.42 &	40.0 &	0.32&	0.16&6.5\\
1600-5044	& 2 & 11 &	0.19&	-136.47 &	6.0 &	0.40&	1.38&4.2\\
1645-0317	& 5 & 6 &	0.39&	-11.85 &	34.5 &	0.36&	0.44&7.4\\
1709-1640   & 4 & 4 &	0.65&	-14.80	&	16.4 &	0.40&	0.39&8.0\\
1731-4744	 & 8 & 9 &  0.78&	-237.62		 &	0.8& 0.40&	0.66	&3.7\\
1752-2806	& 12 & 8 &	0.56&	-25.69		 &	11.0&0.20&	0.29	&8.3\\
1825-0935	& 3 & 3 &	0.77&	-88.78		 &	2.3 &0.27&	0.14	&8.2\\
1932+1059	& 4 & 9 &	0.23&	-22.56		 &	31.0 &0.31&	0.20	&8.3\\

\hline 
\end{tabular}
\end{center}
Note: column two and three denote the number of microglitches of positive and negative signature respectively, column four, five, six denote the rotational period, spin-down rate and characteristics age of the corresponding pulsar respectively, column, seven and eight denote the rate of microglitches of positive and negative signature respectively, while column nine denote the distance from the galactic center.
\end{table*}
\begin{figure}
\centering
\includegraphics[scale=0.7]{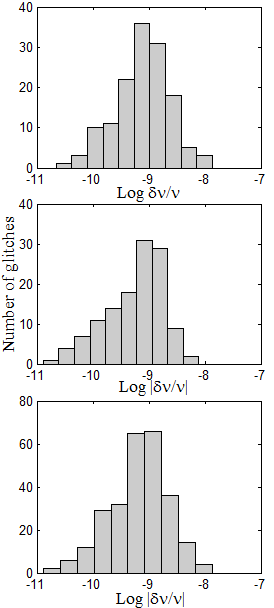}
\caption{Distribution of microglitches: top panel is distribution of glitch sizes of positive sign ($ \delta\nu/\nu $), middle panel is distribution of glitch sizes of negative sign ($ -\delta\nu/\nu $), bottom panel is the combined distribution of glitch sizes of both signs.}
\end{figure}

\section{Starquake Model}
The standard starquake models of pulsar glitches are generally done by parametrizing the dynamics of the neutron star solid crust and core \citep{bru,b5a}.
As a pulsar spins down due to electromagnetic braking torque, strain energy develop until the star can no longer withstand the load on it due to stress.
This leads to a starquake that relives the star of the stress. 
\cite{bru} who was the first to treat pulsar glitch in the realm of starquake 
proposed that the entire strain energy accumulated in the star is released in the quake, while
\cite{b5a} argued that only a part of the strain energy is released as the star is not relieved completely of the stress and the plastic flow is negligible.
Later \cite{b32} in their model presented a situation of which during the quake, the entire stress is almost relived at once as the quake cracks the neutron star.
The total energy released is converted to kinetic energy of the plastic flow and thermal energy.
Some part of the kinetic energy, due to anelastic flow is restored back.
Therefore, the neutron star retains a part of the strain energy, which could partake in subsequent glitch.
However, in these models, the quake only occurs when the neutron star could no longer sustain the stress on it.
Here, we present a situation in which a glitch could occur in neutron star without being stressed to a critical limit.
In this situation, a glitch could occur if the neutron star accrete another body. 
The signature of the glitch depends on the impact of the interaction.
If the impact is high enough to cause quake, microglitch of positive signature is expected else otherwise.
The size of the glitch could at most be equal to that of conventional starquake depending on the stressed state of the neutron star. 

\section{Strange nuggets and microglitch}
Pulsars in the vicinity of strange nuggets could collide with it or even accrete a clump of strange nuggets.
The resulting situation could be highly complicated. 
There could be burst and the spin rate of the pulsar will be affected. 
The latter is the one that could easily be noticed in pulsar timing data. 
During the accretion, the effect in a given pulsar rotational kinetic energy (or spin frequency) depends on the impact of the interaction.
Though it is quite challenging to quantify the total energy involved in the process, coupled with the fact the full nature of the SN is not yet well understood, it can be estimated by calculating the change in energy of the pulsar involved.
At the onset of the interaction, the total energy of the pulsar is expected to change --- observed as a sudden change in kinetic energy (via spin frequency) of the pulsar. 
The effect could be visualised as follows:

\subsection{Case one}
If the impact is strong enough to cause a quake.
Following the standard starquake model, the change in stellar moment of inertia ($ I =   2/5$ $MR ^{2} $) is 
\begin{equation}
\delta I  =  \frac{4}{5} MR \delta R,
\end{equation}
and conservation of angular momentum ($L = 2\pi I\nu $) of the pulsar entails
\begin{equation}
\delta L  = 2\pi [I \delta \nu + \nu \delta I] = 0,
\end{equation}
leading to
\begin{equation}
\frac{\delta \nu}{\nu} = -\frac{\delta I}{I} = - \frac{2\delta R}{R},
\end{equation}
implying that the star spins up to compensate for the change in stellar moment of inertia, where $ \nu $ is the spin frequency and $ \delta\nu/\nu $ is the observed glitch size.
The energy involved in the above process can be obtained following \cite{b5a,b32,b31}.

A rotating neutron star is more oblate in shape than spherical. 
The oblateness, which is a measure of the deviation of the gravitation energy with respect to the non-rotating case is of the form 
\begin{equation}
\epsilon = \frac{I-I_{0}}{I_{0}}, 
\end{equation}
while the eccentricity is
\begin{equation}
e^{2} = 1 - \frac{c^{2}}{a^{2}},
\end{equation}
where $ I_{0}$, $ a $ and $ c $ are the stellar moment of inertia in a non-rotating state, semi-major and semi-minor axes respectively. 
For slow rotators, the oblateness is related to the eccentricity as \citep{b31} 
\begin{equation}
\epsilon = \frac{1}{(1-e^{2})^{1/3}} - 1 \approx \frac{e^{2}}{3}
\end{equation}
and with angular frequency ($ \Omega = 2\pi/P $) as \citep{b31}
\begin{equation}
\Omega = 2e\sqrt{\frac{2\pi G\rho}{15}},
\end{equation}
where $ P $ is the rotation period, $ G $ the universal gravitational constant and $ \rho $ the stellar density.
So that
\begin{equation}
\epsilon \approx \frac{8.83}{3P^{2}} \times 10^{-7}.
\end{equation}
This implies that for rotational period of not less than 0.1 s just like those of the pulsars in this analysis, $ \epsilon \approx 2.94 \times 10^{-5} $. 
Then the change of oblateness with the change in stellar moment of inertia during the glitch is of the form,
\begin{equation}
\delta\epsilon = \frac{\delta I}{I}(\epsilon + 1) \approx \frac{\delta I}{I},
\end{equation}
as $ (\epsilon \ll 1) $.\\ 
The rotational kinetic energy of the star is
\begin{equation}
E_{rot} = \frac{L^{2}}{2I} = \frac{L^{2}}{2I_{0}(\epsilon + 1)}.
\end{equation} 
As the star is a self gravitating body, the gravitational energy is of the form

\begin{equation}
E_{g} = E_{0} + A\epsilon^{2},
\end{equation}
where $ E_{0} $ is the energy in a non-rotating state, $ A = \frac{3}{25} GM^{2}/R $ is a constant for an incompressible sphere, while the elastic energy is of the form 

\begin{equation}
E_{el} = B(\epsilon - \epsilon_{0})^{2},
\end{equation}
where $B = \mu V/2 $, is a constant, $ \mu $ is the shear modulus and $ V = \frac{4}{3} \pi R^{3} $ is the volume of the star, $ \epsilon_{_{0}} $ is the reference oblateness (unstressed star\footnote{If the star is unstressed, $\epsilon = \epsilon_{0}$ and the elastic energy is zero.}).
Then the total energy of such a pulsar is of the form
\begin{equation}
E_{T} = \frac{L^{2}}{2I} + E_{0} + A\epsilon^{2} + B(\epsilon-\epsilon_{0})^{2}.
\end{equation} 
Each glitch is seen as a sudden change in the rotational kinetic energy (via change in spin, $ \delta\nu $) of the pulsar.
In the starquake model, this change is due to change in the oblateness of the star ($ \delta\epsilon $). 
As such, at any interval the star is in equilibrium, it follows that \citep{b31}
\begin{equation}
\frac{\delta E_{_{T}}}{\delta \epsilon} = 0.
\end{equation}
Now, assuming a given pulsar is in equilibrium at a hypothetical time $ t_{1} $ with angular momentum $ L_{1} $, applying the equilibrium condition gives
\begin{equation}
\frac{\delta E_{T}}{\delta \epsilon} =  2A\epsilon_{_{1}} + 2B(\epsilon_{1} - \epsilon_{0}) - \frac{L^{2}_{_{1}}}{2I_{0}(\epsilon_{_{1}} + 1)^{2}}  = 0.
\end{equation}
Now as the pulsar spin-down, elastic energy develops, at the onset of a glitch at angular momentum $ L_{2} $ after an interval $ t_{2} $, applying equilibrium condition gives,

\begin{equation}
\frac{\delta E_{T}}{\delta \epsilon} =  2A\epsilon_{_{2}} + 2B(\epsilon_{_{2}}-\epsilon_{_{0}}) - \frac{L^{2}_{_{2}}}{2I_{0}(\epsilon_{_{2}} + 1)^{2}} = 0,
\end{equation}
In order to estimate the magnitude of elastic energy at the interval, $ \delta t = t_{2} - t_{1} $, the difference of equation (15) and (16) is obtain, thus,

\begin{equation}
2(A + B) (\epsilon_{_{1}} - \epsilon_{_{2}}) = \frac{L^{2}_{_{1}}}{2I_{0}(\epsilon_{_{1}} + 1)^{2}} - \frac{L^{2}_{_{2}}}{2I_{0}(\epsilon_{_{2}} + 1)^{2}}.
\end{equation}
Making use of $ \epsilon + 1 = \frac{I}{I_{0}}$ and $L = 2\pi I\nu $, the change in the 
two equilibrium reference point during spin-down period $ \delta t $ is 
\begin{equation}
(\epsilon_{_{1}} - \epsilon_{_{2}}) = \frac{1}{2(A+B)}[2\pi^{2} I_{0}(\nu^{2}_{_{1}}-\nu^{2}_{_{2}})].
\end{equation}
The expression in the square bracket is the change in the rotational kinetic energy, $ -\delta E_{k} $.
If at any point during the spin-down phase, the pulsar accrete SN, and the impact is high enough to cause a quake, elastic energy is released.
Though it could be difficult to constrain the magnitude of the elastic energy accumulated during the spin-down phase $ \delta t $ it should be proportional to the change in rotational kinetic energy of the pulsar and change in the stellar oblateness.
This energy is at most equal to that due to normal starquake.
Thus:
\begin{equation}
E_{els} \leq B(\epsilon_{_{1}} - \epsilon_{_{2}})^{2} = \frac{B(\epsilon_{_{1}} - \epsilon_{_{2}})}{2(A+B)} (|\delta E_{k}|).
\end{equation}
$\epsilon_{_{1}} $ and $\epsilon_{_{2}} $ in Equations (15) and (16) respectively are two different points at which the star is in equilibrium (unstressed).
This implies that ($\epsilon_{_{1}} - \epsilon_{_{2}} $) is just the change in reference oblate positions $ \delta\epsilon_{_{0}} $ during the normal spin-down phase at the interval $ \delta t $. 
Then to relate it with oblateness change, $ \delta\epsilon $, during glitch, $ -\delta\nu/\nu $, we follow \cite{b32} and apply the equilibrium condition to Equation (13) and making use of $ (\epsilon + 1) = \frac{I}{I_{0}}$ and $ L = 2\pi I\nu $, one readily obtain
\begin{equation}
\epsilon_{_{0}} = \frac{(A + B)}{B}\epsilon - \frac{\pi^{2}I_{0}\nu^{2}}{B}.
\end{equation} 
Making use of Equations (3) and (9), this implies that
\begin{equation}
\delta\epsilon_{_{0}} = -\frac{(A + B)}{B}\delta\epsilon = \frac{(A + B)}{B}\left(\frac{\delta \nu}{\nu}\right).
\end{equation}
Then the total energy released in Equation (19) as result of glitch of size $ \delta \nu/\nu $ is
\begin{equation}
E_{rel} \leq -\frac{1}{2} (|\delta E_{k}|)\left(\frac{\delta \nu}{\nu}\right),
\end{equation}
This energy is seen as enhanced luminosity of the star at the onset of the glitch, 
which can be written as is 
\begin{equation}
\delta E_{rel} \leq -\frac{1}{2}\left[4\pi^{2}I\nu \dot{\nu}\,\,\right] \delta t \,\,\left(\frac{\delta \nu}{\nu}\right) = \frac{E_{k}}{2}\frac{\delta t}{\tau_{c}}\left(\frac{\delta \nu}{\nu}\right),
\end{equation} 
where $ \delta t $ is the inter-glitch time interval, while $ \tau_{c}$ is the characteristic age of the pulsar.
On average, pulsars in this analysis spin-up at intervals of $ \sim 6.3 \times 10^{7} s$.
As such for their typical period of P $ \gtrsim $ 0.1 s and considering that they are conventional pulsars\footnote{i.e. spin-powered, M = 1.4 $ M_{\odot} $ and R = 10 km}, 
\begin{equation*}
\delta E_{rel} \lesssim 1.2 \times 10^{31}\,erg\,\left(\frac{I}{10^{45}gcm^{2}}\right)\left(\frac{0.1s}{P}\right)\left(\frac{-\dot{\nu}}{10^{-15}}\right)
\end{equation*}
\begin{equation}
\qquad\qquad\qquad\qquad\qquad\qquad  \left(\frac{\delta t}{6.3\times 10^{7} s}\right)\left(\frac{\delta\nu/\nu}{10^{-9}}\right).
\end{equation}
In \cite{b8b}, the SN-triggered glitches is taken to be $ \delta\nu/\nu < \delta\nu/\nu(E_{rel} = 10^{35}) $erg.
Considering that and comparing it with the limit of energy in Equation (24), it is reasonable to say that the sizes of the microglitches of positive signatures in this analysis are well accommodated in the range suggested to have a possible link with SNs.   
On the other hand, comparing this amount of energy with the upper limit of glitch energy estimated from X-ray flux enhancement after 35 day Velar pulsar glitch $E_{re} \approx  3.6 \times 10^{36}$ erg \citep{bhel} indicates that this amount is small to produce any detectable burst.
This is also well understood; energy released in conventional starquake is inversely proportional to the pulsars characteristic age.
Velar pulsar is far younger than pulsars in this analysis.

\subsection{Case two}
If the motion of the SN is tangentially opposite to that of the pulsar, at the time of the interaction, the impulse could manifest in sudden change in stellar angular momentum. 
The resultant rotational kinetic energy of the system is,
\begin{equation}
\frac{L^{2}}{2I} - E_{_{SN}} = \frac{L^{2}}{2I} + \frac{L\delta L}{I} + \phi,
\end{equation}
where the second term in the right-hand-side of Equation (25) is the change in rotational kinetic energy and $ \phi $ accounts for other form of energy loss, which could be burst. 
This implies, 
\begin{equation}
E_{_{SN}} = - \frac{L\delta L}{I} + \phi.
\end{equation} 
The ratio of the SN energy to that of pulsar rotational kinetic energy gives, bearing in mind that is negligible, 
 \begin{equation}
\frac{E_{_{SN}}}{E_{rot}} = - 2\frac{\delta L}{L} = - 2\frac{\delta \nu}{\nu},
\end{equation}
which reduces to,
\begin{equation}
E_{_{SN}} = \frac{4\pi I}{P^{2}} \left(-\frac{\delta \nu}{\nu}\right).
\end{equation}
If the energy of SN or its clump satisfies Equation (26), microglitch of negative signature occur. 
The occurrence rate now depends on the position of the pulsar relative to the galactic center, as we discuss latter.

\section{Energy of Strange nuggets}
The actual nature of strange nuggets is still not well understood, but it is believed to interact gravitational as dark matter circulating in the galaxy. 
They can be orbiting a more massive and magnetised objects like neutron star within the galaxy. 
They can even be in isolation or interact with each others that the chances of existing in clumps could be there, just like multiple planets in a system colliding with each other, produce some clumps with negative angular momentum \citep{katz1994,ford2008,bhg}. 
As such the energy range of SN could be wide depending on the form of a given one.
However, for simplicity, we can still treat the energy of SNs as that of a single object and that of the hypothetical clumps to be an integral sum of each.
 
For dark matter in form of SN to be in line with the prediction of Big Bang Nucleonsynthesis and the observed abundance of helium, it is expected that the baryon number ($ A $) of each SN should be $ A \geq 10^{24} $ \citep{b9a}.
On the other hand, if the formation of high red-shift ($ Z \approx 6 $), supper massive black-holes ($ 10^{9} M_{\odot} $) are the consequences of SNs \citep{b7F}, it suggests that $ A \leq 10^{35} $ \citep{b8aa,b8b}.
Therefore, one can constrain the baryon number of SNs in the range $ 10^{24} \leq A \leq 10^{35} $. 
However, for those existing in clumps, it could be more than this.

Now for a typical isolated SN of mass $ m_{_{SN}} = A m_{_{b}}$ approaching from outside the gravitational field of a given pulsar, where $ m_{_{b}} = 1 \,G \,eV$ is the baryon mass \citep{b8b}, just before it hits the neutron star surface, its velocity is
\begin{equation}
V_{_{SN}} = \left(\frac{2GM}{R}\right)^{1/2}.
\end{equation}
The energy as it hits the stellar surface is 
\begin{equation}
E_{_{SN}} = \frac{1}{2} m_{_{SN}}V^{2}_{_{SN}} = \frac{GMm_{_{SN}}}{R}= 3.3\, A \times 10^{-4}\,erg.
\end{equation}
Therefore, if a SN is accreted by pulsar, it's energy depends on gravitational pull of the pulsar and the mass of the SN.
However, if the SNs exist in clumps, the total energy is hard to constrain but it is believed to be more than the one above.

On accretion of SN, the structure of the neutron star will be affected as their could be interaction of SN with particles inside the neutron star crust. 
The details of such interaction is beyond the scope of this work, coupled with the fact that the actual nature of SN is not well understood. However, following \cite{b32,b8b}, and believing that neutron stars have solid crust that can sustain stress, a consequence of such interaction could be estimated.
As the neutron star spins down, stress builds in the crust, at a critical $ \sigma_{c}$, ($ \epsilon - \epsilon_{0} $) $ \sim 10^{-3} $ beyond, which the crust crack, 
the strain energy density is
\begin{equation*}
E_{_{\rho}} = \frac{E_{el}}{V} = \frac{1}{2}\mu \left(\frac{\epsilon - \epsilon_{0}}{10^{-3}}\right) = 
\end{equation*}
\begin{equation}
\qquad \frac{1}{2}\times 10^{26}ergcm^{-3}\left(\frac{\mu}{10^{32}ergcm^{3}}\right)
\left(\frac{\epsilon - \epsilon_{0}}{10^{-3}}\right).
\end{equation}
The lattice density inside a neutron star is $\sim 0.3 \,fm^{-3}$ ($ 0.3 \times 10^{39} \,cm^{-3} $), then to brake the lattice structure, each of the lattices should absorb $1.67 \times 10^{-13} erg $.
If each of the lattice receives $ L_{_{E}} $ of energy, the total number of lattice it shall interact with is 
\begin{equation}
T_{_{N}} = 2.0 \times 10^{9}\,A\left(\frac{1.67\times 10^{-13}\,erg}{L_{_{E}}}\right).
\end{equation}
The interaction cross section is at most that of the SN.
Now assuming that the interior density of SN is comparable to that of dark matter, it cross section is $ \sim \pi(A\,\frac{3}{4\pi} \times 10^{-39})^{2/3}\,cm^{2}$.
For the trajectory of the interaction to traverse the entire length of the neutron star radius, it will interact with $ N_{_{L}} $ number of lattices in a volume element $ \sim \pi(A\,\frac{3}{4\pi} \times 10^{-39})^{2/3}\,cm^{2}\, \times 10^{6}\,cm$.
Thus the number of lattices in this volume element is 
\begin{equation}
T_{_{int}} = \left[\pi\left(A\frac{3}{4\pi} \times 10^{-39}\right)^{2/3}\times 10^{6}\right]\,\times 0.3\times\,10^{39}.
\end{equation}  
Therefore, if $T_{_{N}} \geq T_{_{int}} $, an induced quake could occur.
The minimum energy required of the SN is the quantity of energy absorbed by each of the lattice structure multiple by $ T_{_{int}} $.
This implies that for SN of $ A \geq 10^{32} $, a microglitch of positive signature is feasible. 
However, this depends on the line of action of the interaction. 
If it is perpendicular to the neutron star surface, the possibility of the length of the trajectory reaching the neutron star core is there. 
Therefore, it entails that with single SN of $A > 10^{32}$, of which its line of action is perpendicular, the interaction could lead to a quake.
On the other way, if a neutron star accrete a clump of SNs and the line of action is not perpendicular to the surface, and the trajectory is not towards the core, the neutron star could survive induced quake. 
But the impulse could translate to an observable sudden change in angular momentum of the pulsar, if the energy of the strange nuggets clump is high enough.

\section{Microglitch rate and accretion rate of strange nuggets}
To work out the expected glitch rate, the rate of oblateness change during the glitch ($ -\delta\nu/\nu $) and that during the normal spin-down phase ($ \epsilon_{1} - \epsilon_{2} $) are considered.
From Equation (18), the rate the oblateness is changing during the normal spin-down phase is
\begin{equation}
\dot{\epsilon}_{_{0}} = \frac{2\pi^{2}I_{0}}{(A + B)}\left(\frac{-\dot{\nu}}{P}\right) \sim 10^{-13}\,yr^{-1}\left(\frac{-\dot{\nu}}{10^{-15}}\right)\left(\frac{0.1\,s}{P}\right),
\end{equation}
for a mean share modulus $ \mu = 10^{32}\,$erg cm$^{3} $.
This is related to that during the glitch via Equation (20) as
\begin{equation}
\dot{\epsilon}_{_{0}} = \left(\frac{A + B}{B}\right)\dot{\epsilon} \sim 10^{2}\dot{\epsilon}.
\end{equation}
These rates are expected to be constant in a given pulsar for given values of $ \dot{\nu} $ and $ P $ (which are the parameters that quantifies the characteristic ages of the pulsars) as A, B and I$_{0}$ are expected to be constants.
This implies that
\begin{equation}
\dot{\epsilon} \sim 10^{-11}\, yr^{-1} \left(\frac{-\dot{\nu}}{10^{-15}}\right)\left(\frac{0.1\,s}{P}\right),
\end{equation}
from, which it is readily ascertain that 
\begin{equation}
\delta\,t \sim 10^{2}\, yr\left(\frac{\delta\nu/\nu}{10^{-9}}\right)\left(\frac{10^{-15}}{-\dot{\nu}}\right)\left(\frac{P}{0.1\,s}\right),
\end{equation}
using the modal microglitch size in Figure (1).
For a typical size of 10$^{-9}$, it will take an ideal pulsar of period 0.1 s, spin-down rate 10$^{15}$ Hzs$^{-1}$ an interval of $ \sim $ 100 yr to suffer starquake.
Considering the pulsars in this analysis for such a glitch, it will take between tens to thousands of years for quake of microglitch size of positive signature to occur. 
Therefore, the standard starquake model is insufficient to explain the rate of microglitches of positive signature. 
However, the possibility of induced starquakes is not ruled out.

To compare the microglitch rate with the accreation rate of dark matter by pulsars at a distance $ r $, from the galactic center, we follow \cite{b8b} believing that the distribution of SN in the galaxy is similar to that of dark matter and the mass fraction is $ \gamma = 0.1$.
The accretion rate of SNs by pulsar with mass M and radius R is \citep{Press,Kouvaris}.
\begin{equation}
\textit{f} = 4\pi^{2}n_{_{x}}\left(\frac{3}{2\pi v^{2}}\right)^{3/2} \,\, \frac{2GMR\, v^{2}}{3\Lambda},
\end{equation}
where $ \Lambda = (1- 2GM/R)^{-1}$ is a factor accounting for general relativity effect,
$n_{x} = \gamma.\rho_{_{dm}}/m_{_{SN}}$ is the number density of the SNs, $ m_{_{SN}} = Am_{b} $ is SN mass and $ M_{b} \sim 1 GeV $ is baryon mass.
\begin{equation}
\rho_{_{dm}}(r) = \frac{\rho_{_{s}}}{(\frac{r}{r_{_{s}}})\langle1 + (\frac{r}{r_{_{s}}})^{2}\rangle}
\end{equation}
 is the dark matter distribution as a function of distance from the galactic center (r) \citep{Navarro}. 
 $ \rho_{_{s}} = 0.26\,GeV\, cm^{-3}$ and $ r = 20$ kpc . 
 The circular velocity of the SN is of the form
 \begin{equation}
 v(r) = \sqrt{\frac{GM(r)}{r}},
 \end{equation}
where $ M(r) = \int_{_{0}}^{^{r}}\rho(r ^{\prime})4\pi r^{\prime2}\,dr^{\prime} $ is the total within a radius r.

\begin{figure}
\centering
\includegraphics[scale=0.75]{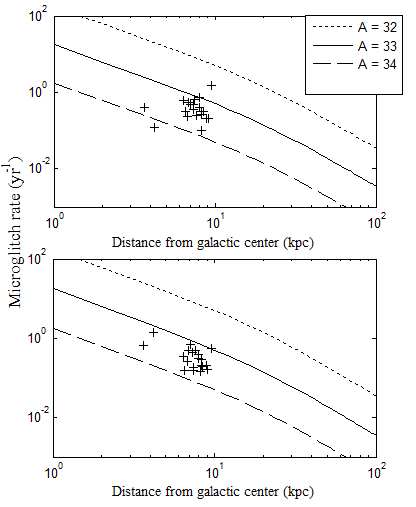}
\caption{Accretion rate of SN by pulsar at a distance r from the galactic center compare with the microglitch rate. Top panel is for microglitch of positive signature, bottom panel is for microglitch of negative signature.}
\end{figure}

The microglitch rate as compared with accretion rate of SNs by pulsars is shown in Figure (2). 
Such a figure has earlier been presented by \citep{b8b}.
From Figure (2), the pulsars are relatively at distance of which the interactions with SN are feasible. 
The lines represent the ideal accretion rate of SNs of a given baryon number by pulsars at a distance r from the center of the galaxy, while the data points, represent the microglitch rate of a pulsar at distance r from the center of the galaxy.

\section{Discussion}
In this analysis, we propose that microglitches could be consequences of accretion of SNs by pulsars.
These SNs are believed to be the relics of quantum chromodynamic phase transition that took place at the earliest stage of the universe evolution.
As such, if they survived the present universe, they could fill our galaxy and could be colliding with (or accreted by) stellar objects.
Pulsars due to their high gravitational field are good candidates for such accretion especially pulsars of relatively long period \citep{b6b}.
In addition, due to kick velocity of pulsars, they are bound to collide with matter along their path and such collisions should have astrophysical consequences of which sudden change in rotational kinetic could be one of them.
In these frameworks, the interaction of pulsars with SNs could lead to starquake (or cracking of the crust) if the impact is high enough. 
The quake readily lead to sudden spin-ups observed as microglitches of positive signatures.
In contrast, if the energy of the SNs is not large enough for the impact to cause starquake, the interaction leads to sudden increase in stellar moment of inertia, which manifests in sudden spin-down observed as microglitches of negative signatures. 
Thermal X-ray burst could readily accompany these kind of glitches, but no such observation was reported.    
This could be that the observers (of the microglitches) did not consider that in their analysis, or that it could not be detected due to low magnetic field of the pulsars involved.
Glitches, which have been associated with X-ray bursts or other radiative changes, are mostly seen in magnetars and high magnetic field pulsars \citep [e.g.][]{bwje,bdk,bka,bkou}.
If this model is to be extended to such pulsars, it could be a good tool for understanding some of the radiative changes accompanying their glitches.
This is as the interaction of SN with high magnetic field pulsars could be a complicated process.
There could be tidal heating, the SN may disintegrate of which part of it may be ionized.
The ionized matter could readily affect the pulsar profile (radiative change) as they interact with the pulsar magnetosphere. 

Figure 2 compare the accretion rates of SNs of baryon number A by pulsars at a distance r from the galactic center with the microglitch rate of pulsars at the same distance. 
It is readily observed that the positions of the pulsars are within the location that they can accrete SN. 
It appears that SN of A $> 10^{32} $ is more likely to interact with pulsars at such distances. 
A striking feature is that the pulsars concentrate highly within that region, indicating that they are all prone to a phenomenon associated with that region.  
Nonetheless, the narrow dispersion in the distribution of the pulsar distances makes it quite challenging to reasonably analysis the correlation between the microglitch rates and pulsar distances from the galactic center, though, a Pearson correlation, yielded at least a significant result. 
That is, the rate of microglitches of negative signature, correlating well ($r = - 0.61$, Figure 2 middle panel) with pulsar distances from the galactic center. 
This is an indication that interaction with SNs can lead to microglitches especially those with negative signatures.
Pulsars closer to the galactic center accretes SNs more often than those farther, leading to more sudden spin-downs in the pulsars than in others.
This is in line with recent analysis that suggested that anti-glitch (glitch with negative signature) in magnetar resulted from collision of the magnetar with solid body \citep{bhg}. 
Equally, it is believed that as pulsars gets older, they migrate towards the galactic center,  
so regions nearer to the galactic center should have higher concentration of older pulsars than others.
Therefore, microglitches being more prominent in older pulsars should be understood in that realm. 
This could also explain the narrow dispersion in the distribution of pulsar distances involved in this analysis as they are of similar characteristic age.   
The weak correlation of microglitches of positive signature with pulsar distances from the galactic center could be that, some the glitches could be from conventional starquake, as there is no criterion that inhibits a crust that has been stressed beyond the limit from cracking.

On the other hand, pulsars involved in this microglitch activities are relatively old and low magnetic field objects, which are not spinning down fast to warrant the glitch rate they posses. 
Considering, their spin-down rate ($ \dot{\nu} $) and their rotational periods, the inter-glitch time intervals in most them exceeds 100 yrs, unlike the $ \sim 2yr $ inter-glitch time interval seen them. 
Small size glitches of such rate are mostly seen in younger pulsars of high spin-down rate \citep{b7}. 
It is argued that in those young pulsars, due to the higher temperature in their interior, vortex unpinning may not work efficiently as thermal fluctuations could in fact contend against pinning forces and hamper the creation of large pinning zone for the vortices \citep{b7}, 
but not for older pulsars, which are believed to have cooler interiors.
Equally, lesser spin-down rates would entail larger time intervals between glitches (i.e. low glitch rate), not essentially a reduction in glitch sizes, as the glitch size depends directly on the number of unpinned vortices \citep{bwm} and inter-glitch time intervals are size independent \citep{b7e2}. 

In conclusion it is clear that the negative signature of some microglitches are not in line with the standard glitch models. 
This paper has presented a way of understanding the glitches based on a possible natural phenomenon --- the interaction of dark matter inform of strange nuggets with pulsars.




\nocite{*}
\bibliographystyle{spr-mp-nameyear-cnd}
\bibliography{biblio-u1}


\end{document}